\documentclass[12pt]{article}
\usepackage{epsfig}

\begin{document}

\thispagestyle{empty}
\renewcommand{\thefootnote}{\fnsymbol{footnote}}

\begin{flushright}
{\small
SLAC--PUB--8775\\
February 2001\\}
\end{flushright}

\vspace{.8cm}

\begin{center}
{\bf\large   
A study of topological vertexing for heavy quark 
tagging\footnote{Work supported by
Department of Energy contract  DE--AC03--76SF00515.}}

\vspace{1cm}

Toshinori Abe\\
Stanford Linear Accelerator Center, Stanford University,
Stanford, CA  94309\\

\end{center}

\vfill

\begin{center}
{\bf\large   
Abstract }
\end{center}

\begin{quote}
We compare heavy quark tagging and anti-tagging efficiencies for 
vertex detectors with different inner raddi
using the topological vertex technique developed at the SLC/SLD experiment.
Charm tagging benefits by going to very small inner radii.
\end{quote}

\vfill

\begin{center} 
{\it Talk presented at} 
{\it 5th International Linear Collider Workshop (LCWS 2000)} \\
{\it Fermilab, Batavia, Illinois} \\
{\it 24-28 Oct 2000} \\ 

\end{center}

\newpage



%
\pagestyle{plain}

\section{Introduction}

A vertex detector (VTX) is a very poweful particle identification device 
for the future linear collider experiment.
VTX allows not only $b$/$c$-jet tagging but also anti-$b$/$c$ jet tagging.
Excellent $b$/$c$-jet tagging is required in studies of  Higgs and Top physics.
VTX performance depends critically on the innermost radius ($r_{inner}$) 
of the detector.
Many studies have been done to achieve smaller $r_{inner}$ 
in order to get better impact parameter resolution.
Current allowable $r_{inner}$ is expected to be $\sim 1$~cm.
However a VTX configuration with $r_{inner}=1$~cm is very difficult to achieve
and there are presently no physics studies comparing
such aggressive designs with more conservative ones.
In this paper, we discuss the physics-performance difference 
between $r_{inner}=1$~cm and 2~cm VTX configurations,
using the topological vertexing technique developed at the SLC/SLD experiment.

\section{Tools}
We use the LCD fast simulation and a topological vertexing and a mass tag 
technique for the study.

The LCD fast simulation~\cite{Iwasaki:2001iq} is based on 
the ROOT analysis tool~\cite{Brun:1997pa} 
and the C++ programming language 
to maximally benefit from object oriented programming techniques.
In this simulation, track particles are smeared according to
their error matrices.
The error matrices are given by a look-up table method based on momentum and 
$\cos\theta$ of charged particles.
The error matrices include off-diagonal elements to give added realism.
The vertex detector is assumed to have layers at several radii 
($r$ = 2.4~cm, 3.6~cm, 4.8~cm, 6.0~cm) and resolution of 5~$\mu$m
for each layer in both detector configuration.
The VTX configuration with $r_{inner}=1$~cm has an extra layer of $r$ = 1.2~cm
with resolution of 5~$\mu$m.

The success of the CCD-based VTX at the SLC/SLD 
experiment~\cite{Abe:1997bu,Abe:2000ky}
argues strongly that a CCD-based VTX will provide optimal performance
in a future linear collider experiment.
Taking advantage of the precise 3-D spatial points
provided by the VTX, 
a topological vertexing technique~\cite{Jackson:1997sy}
has been developed.
The topological vertexing naturally associates tracks
with the vertices where they originated and
can reconstruct a full $b$/$c$-meson decay
chain, i.e, primary, secondary, and tertiary vertices.
Using the reconstructed secondary/tertiary vertex, 
the invariant mass of the tracks associated with decay
is used to identify jet flavor (mass tag technique~\cite{Abe:1998sb}).
This combination of the techniques gives the best heavy-flavor-jet tagging
performance in $e^+e^-$ colliding experiments at present.
Here it should be noted that the secondary/tertiary vertex reconstruction 
enables vertex charge information to be determined 
which gives quark/anti-quark jet identification even 
for neutral $B$'s~\cite{Abe:2000gp}.
The original vertexing program, called ZVTOP, 
was written in Prepmort programing language;
we translated the code into the C++ language in order to suit
the environment of the LCD fast simulation more naturally.
Other physics studies which use the program are reported
in these proceedings\cite{Jim,Iwasaki:2001ip}.

\section{Performance}
In order to investigate the influence of the VTX configuration,
we considered the following variables:
(1) impact parameter resolution;
(2) reconstructed primary vertex resolution; and
(3) $b$-tag, $c$-tag, and anti-$b$/$c$ tag efficiencies and purities.
These studies are done using hadronic decay events at $\sqrt{s}=91.26$ GeV.
The results are summarized in Table~\ref{Tab:compare}.

\begin{table}
\caption{The performance of two different VTX configurations.}
\label{Tab:compare}
\begin{tabular}{l|c c}
\hline
 & $r_{inner}=1$ cm &  $r_{inner}=2$ cm \\ \hline
impact parameter resolution
& $3.2 \mu{\rm m} \bigoplus 8.5 \mu{\rm m} / p \sin^{2/3}$
& $3.5 \mu{\rm m} \bigoplus 14 \mu{\rm m} / p \sin^{2/3}$ \\
reconstructed primary vertex resolution
& $4.6\mu$m($xy$) $3.7\mu$m($rz$) & $6.9\mu$m($xy$) $5.2\mu$m($rz$) \\
$b$-jet tagging efficiency and purity
& $\epsilon=63\%$ $\Pi=97$\%
& $\epsilon=62\%$ $\Pi=97$\% \\
$c$-jet tagging efficiency and purity
& $\epsilon=32\%$ $\Pi=83$\%
& $\epsilon=27\%$ $\Pi=80$\% \\
anti-$b$/$c$ jet tagging efficiency and purity
& $\epsilon=81\%$ $\Pi=91$\%
& $\epsilon=78\%$ $\Pi=90$\% \\
\hline
\end{tabular}
\end{table}

As we expect, $r_{inner}=1$~cm VTX configuration shows better 
impact parameter and
reconstructed primary vertex resolutions than $r_{inner}=2$~cm.
The reconstructed primary vertex resolution, 
in particular $rz$ resolution, is important to 
heavy quark physics at giga-$Z$ experiment.
We also believe that the resolution will play an important role when
we try to discriminate mini-jet backgrounds from Higgs signal 
events~\cite{Matsui}.
This idea needs further study.

\begin{figure} 
\centerline{\epsfig{file=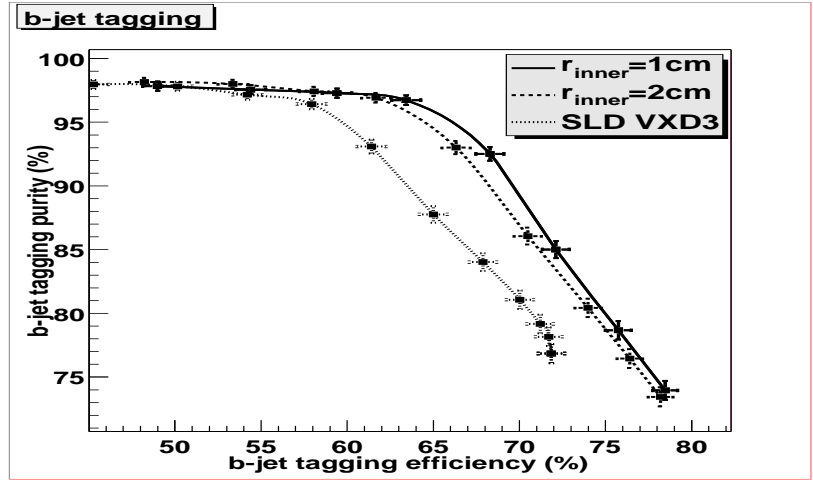,height=2.5in,width=4.2in}}
\caption{Performance of $b$-jet flavor tag.}
\label{Fig:b-tag}
\end{figure}

\begin{figure} 
\centerline{\epsfig{file=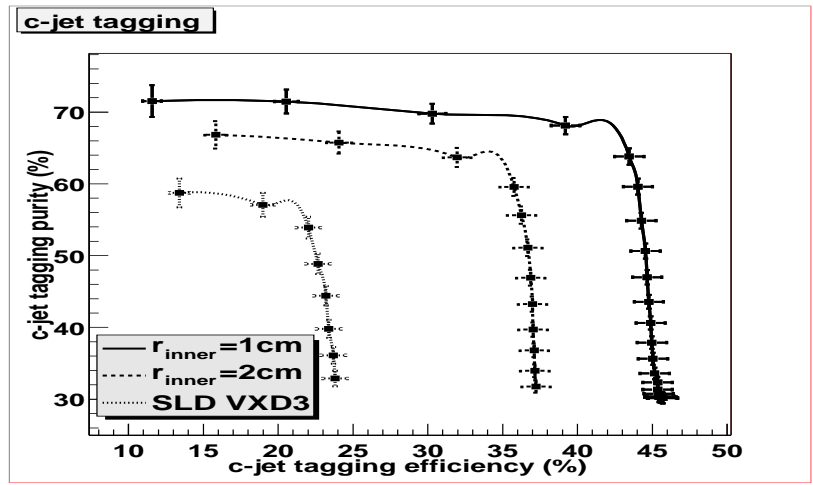,height=2.5in,width=4.2in}}
\caption{Performance of $c$-jet flavor tag.}
\label{Fig:c-tag}
\end{figure}

For jet-flavor identification, we see a result contrary to our
naive expectation.
Figs.~\ref{Fig:b-tag} and \ref{Fig:c-tag} 
show the purity against total efficiency plots for 
$b$-jets and $c$-jets obtained by varying the cut of vertex invariant mass,
respectively.
From these figures,
we can not see significant differences 
in $b$-jet tagging between the two VTX configurations,
but we do see significant improvements for $c$-jet tagging.
This can be understood because the maximum $b$ tag efficiency is
limited by the fraction of decays which ZVTOP can identify,
i.e. those resulting in at least two charged particles.
Furthermore, the long $b$ lifetime ensures that most decays are well-separated
from the primary; hence improved resolution is not needed to find more decay
vertices close to the IP.
With improving VTX resolution the $c$-jet efficiency increases
faster than the $b$-jet efficiency.
We need further study to understand this behavior fully.

\begin{figure} 
\centerline{\epsfig{file=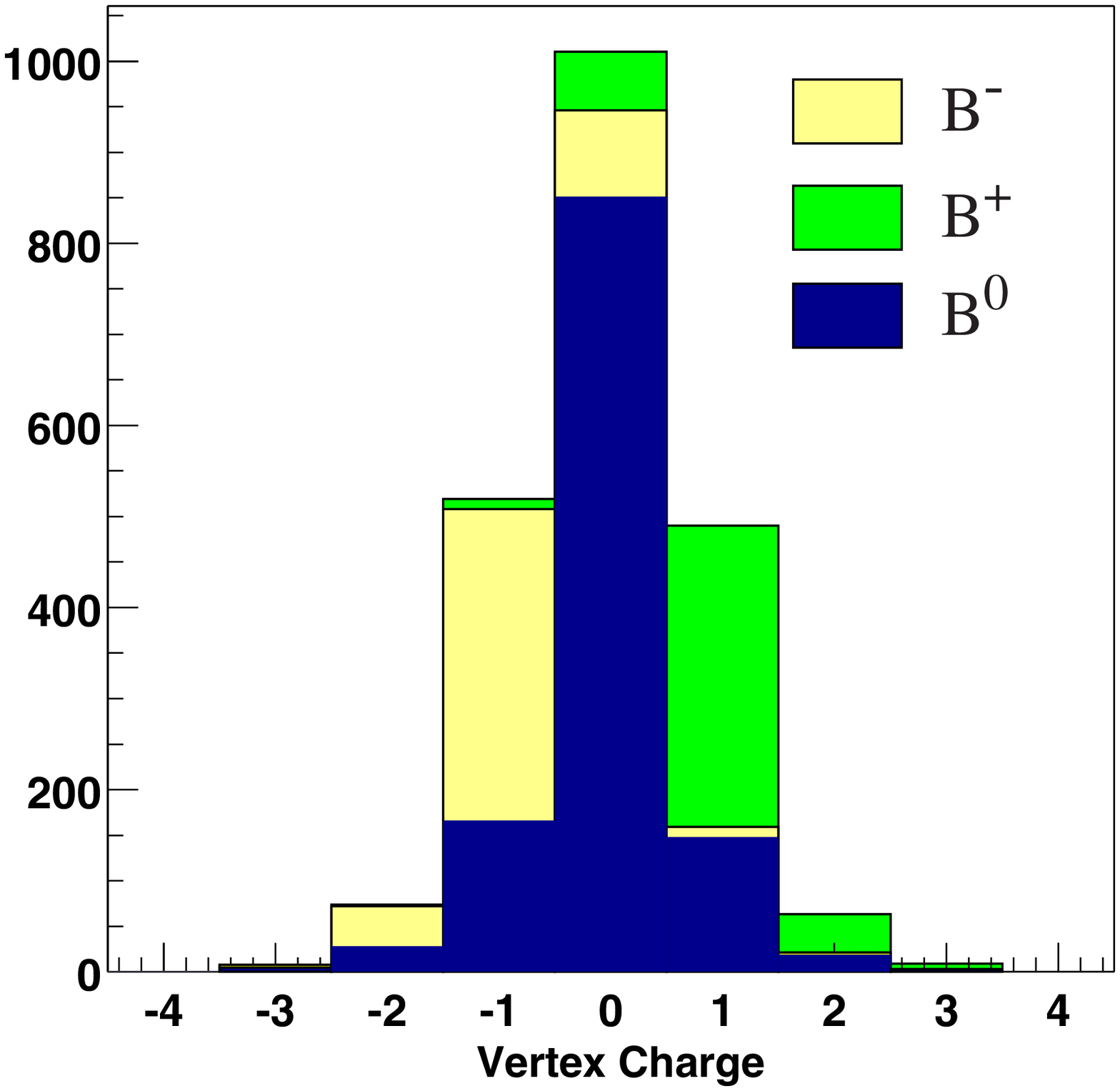,height=2.5in,width=4.2in}}
\caption{Vertex charge.}
\label{Fig:Qvtx}
\end{figure}

In the previous section, we mentioned that the importance of
secondary/tertiary vertex reconstruction.
This is something that has been overheaded in past linear collider studies.
For charged $b$ or $c$ hadrons, vertex charge identifies whether its a quark
or anti-quark jet.
Fig.~\ref{Fig:Qvtx} illustrates the clear charge separation for $B^+$/$B^-$
decay vertex.
According to Ref.~\cite{Iwasaki:2001ip}, 
we can know the $t/\bar{t}$-quark
direction with efficiency and purity of 78~\% and 41~\%, respectively, 
by looking at the charge of the $B$ 
it decays into, requireing $|\cos\theta_{track}|<0.9$.

\section{Summary}
We have developed a fast simulation code to optimize the detector design for
a future linear collider experiment.
First results with a topological
vertexing technique are presented in this proceeding.
The two VTX configurations ($r_{inner}=1$~cm and 2~cm) do not show significant
difference for $b$-jet tagging, but do for $c$-jet tagging.
This should be investigated with further study.

%

 


\begin{thebibliography}{9}

%
\bibitem{Iwasaki:2001iq}
M.~Iwasaki and T.~Abe,
hep-ex/0102015.
%
\bibitem{Brun:1997pa}
R.~Brun and F.~Rademakers,
Nucl.\ Instrum.\ Meth.\ {\bf A389}, 81 (1997).
%
\bibitem{Abe:1997bu}
K.~Abe {\it et al.},
Nucl.\ Instrum.\ Meth.\ {\bf A400}, 287 (1997).
%
\bibitem{Abe:2000ky}
T.~Abe  [SLD Collaboration],
Nucl.\ Instrum.\ Meth.\ {\bf A447}, 90 (2000)
[hep-ex/9909048].
%
\bibitem{Jackson:1997sy}
D.~J.~Jackson,
Nucl.\ Instrum.\ Meth.\ {\bf A388}, 247 (1997).
%
\bibitem{Abe:1998sb}
K.~Abe {\it et al.}  [SLD Collaboration],
Phys.\ Rev.\ Lett.\ {\bf 80}, 660 (1998)
[hep-ex/9708015].
%
\bibitem{Abe:2000gp}
K.~Abe {\it et al.}  [SLD Collaboration],
hep-ex/0012043.
%
\bibitem{Jim} 
J. Brau {\em et al.},
``Higgs Branching Ratio Measurements at a Future Linear Collider'',
to be appeared in the LCWS2000 proceedings.
%
\bibitem{Iwasaki:2001ip}
M.~Iwasaki,
hep-ex/0102014.
%
\bibitem{Matsui}
Prof.~T.~Matsui, private communication.
%

\end{thebibliography}
\end{document}